\documentclass[preprint,aip,superscriptaddress,showpacs]{revtex4-1}

\usepackage{graphicx}
\usepackage{dcolumn}
\usepackage{bm}
\usepackage{natbib}
\usepackage{amsmath}
\usepackage{multirow}
\usepackage{makecell}
\usepackage{hyperref}
\usepackage{bbm}
\usepackage{amsthm,amsmath,amssymb,esint}
\usepackage{mathrsfs}
\usepackage{geometry}
\usepackage{verbatim}
\usepackage{physics}
\usepackage{BOONDOX-calo}
\usepackage{subfiles}
\usepackage{textcomp}
\usepackage{float}

\begin{document}
	\title{Control of energy spectra and enhancement of energy conversion of fast electrons generated by dual-color picosecond lasers}

	\affiliation{Beijing National Laboratory for Condensed Matter Physics, Institute of Physics, CAS, Beijing 100190, China}
	\affiliation{Department of Physics and Beijing Key Laboratory of Opto-electronic Functional Materials and Micro-nano Devices, Renmin University of China, Beijing 100872, China}
	\affiliation{School of Physical Sciences, University of Chinese Academy of Sciences, Beijing 100049, China}
	\affiliation{Key Laboratory of Quantum State Construction and Manipulation (Ministry of Education), Renmin University of China, Beijing, 100872, China}
	\affiliation{IFSA Collaborative Innovation Center, Shanghai Jiao Tong University, Shanghai 200240, China}
	\affiliation{Songshan Lake Materials Laboratory, Dongguan, Guangdong 523808, China}
	\affiliation{Key Laboratory for Laser Plasmas (MoE) and School of Physics and Astronomy, Shanghai Jiao Tong University, Shanghai 200240, China}

	\author{Tie-Huai Zhang}
	\affiliation{Beijing National Laboratory for Condensed Matter Physics, Institute of Physics, CAS, Beijing 100190, China}%
	\affiliation{School of Physical Sciences, University of Chinese Academy of Sciences, Beijing 100049, China}%

	\author{Wei-Min Wang}%
	\email{weiminwang1@ruc.edu.cn}
	\affiliation{Department of Physics and Beijing Key Laboratory of Opto-electronic Functional Materials and Micro-nano Devices, Renmin University of China, Beijing 100872, China}%
	\affiliation{Key Laboratory of Quantum State Construction and Manipulation (Ministry of Education), Renmin University of China, Beijing, 100872, China}
	\affiliation{IFSA Collaborative Innovation Center, Shanghai Jiao Tong University, Shanghai 200240, China}

	\author{Yu-Tong Li}
	\email{ytli@iphy.ac.cn}
	\affiliation{Beijing National Laboratory for Condensed Matter Physics, Institute of Physics, CAS, Beijing 100190, China}%
	\affiliation{School of Physical Sciences, University of Chinese Academy of Sciences, Beijing 100049, China}
	\affiliation{IFSA Collaborative Innovation Center, Shanghai Jiao Tong University, Shanghai 200240, China}
	\affiliation{Songshan Lake Materials Laboratory, Dongguan, Guangdong 523808, China}
	
	\author{Jie Zhang}
	\affiliation{Beijing National Laboratory for Condensed Matter Physics, Institute of Physics, CAS, Beijing 100190, China}
	\affiliation{IFSA Collaborative Innovation Center, Shanghai Jiao Tong University, Shanghai 200240, China}
	\affiliation{Key Laboratory for Laser Plasmas (MoE) and School of Physics and Astronomy, Shanghai Jiao Tong University, Shanghai 200240, China}
	
	\date{\today}
	
	\begin{abstract}
		After the successful fusion ignition at National Ignition Facility, seeking for a high-gain fusion scheme becomes the next hot-spot in inertial confinement fusion community. Fast ignition provides an alternative due to its potential to reduce the energy of driven lasers and achieve higher target gain, whose core step is generating fast electron beam using picosecond lasers. The properties of the electron beam, such as energy spectra, determines the succeed heating process and the yield. In this study, with 2D particle-in-cell simulations, we find that the energy transfer ratio would significantly increase, and the energy spectra would be lifted up in the dual-color injecting scheme, where an extra weak low-frequency laser is injected with the main pulse. These phenomena are attributed to the longitudinal electric field modulation in low-density region and electromagnetically-induced-transparency process in the microchannels formed near the critical surface. Our results can be applied to fast ignition schemes and experimental designs of super-hot electron generation with different pulse durations. For example, in the DCI project with a fast ignition scheme, such a dual-color intense picosecond laser facility is under construction to control the fast electron generation and enhance the laser transparency.
	\end{abstract}
	
	\maketitle
	
	\section{\label{sec:intro}Introduction}
	Since 2021, several breakthroughs have been achieved in inertial confinement fusion (ICF) experiments at National Ignition Facility, marking that the threshold of controlled fusion ignition is approached for the first time \cite{Zylstra2022,Abu-Shawareb2022}. However, the extra energy transfer process at the hohlraum would lead to substantial radiation loss, limiting its steps towards high target gain. Furthermore, there are still problems such as energy conversion efficiency and the balance on compression and hydrodynamic instability \cite{Colaitis2022,Schlossberg2021}. Besides central ignition, fast ignition (FI) \cite{Tabak1994,Kodama2001,Tabak2006,Kitagawa2022} is another possible scheme for ICF, which drives the deuterium–tritium (DT) fuel with weaker nanosecond laser beams compared to the central ignition schemes and uses picosecond pulse(s) to generate super-hot electron beam and ignite the compressed plasma. In this scheme, the energy conversion from laser to super-hot electrons lies in the central position of the whole experiment design. To shorten the transport distance of electron beam and reduce energy losses, a gold cone is widely used in many FI schemes such as cone-in-shell \cite{Nagatomo2019} and double-cone ignition (DCI) \cite{Zhang2020} to keep the ps pulse apart from low-density corona plasma, guiding laser pulse towards the core \cite{Nagatomo2019,Rusby2021}.
	
	The energy spectrum of super-hot electron beam is another key factor in FI scheme, since the electrons have to travel a long distance before reaching the core plasma. The position of energy deposition and hot spot generation are very sensitive to the electron spectrum, which needs detailed optimization for specific schemes. In typical cone-in-shell scheme, this transport distance is around 80 \textmu m \cite{Robinson2014, Nagatomo2021}, which implies that only a part of super-hot electrons lying in the specific energy range can heat the hot-spot region effectively. 
	
	Apart from ICF, the electron beam generated by intense short-pulse (sub-picosecond or femtosecond) laser interacting with solid target also has a wide applications such as generating broadband x-ray emission \cite{Sawada2020} and strong terahertz radiation \cite{Woldegeorgis2019}. There are various temperature scaling laws of super-hot electron generated by laser-plasma interaction (LPI), distinguished from pulse duration and laser intensity \cite{Gibbon1992,Forslund1977,Wilks1992}. When laser intensity reaches the relativistic region, $\bf J\times B$ heating is the dominant acceleration mechanism and the "ponderomotive temperature" [Eq. (\ref{for:temp})] is often used to estimate electron temperature generated by a linear-polarized laser \cite{Wilks1992}, where $I_{18}$ is the intensity in $10^{18}$ W/cm$^2$.
	\begin{equation}
		T_p\simeq 0.511\left[
		\sqrt{1+0.73(I_{18}\lambda_{\mu m}^2)}-1
		\right]\quad (\text{MeV})
		\label{for:temp}
	\end{equation}
	However, the pulse duration and density scale length would strongly affect temperature of super-hot electrons \cite{Ivanov2014,Yabuuchi2010}. Actually, the electron spectrum depends on much more parameters besides laser intensity and wavelength, and the electron energy does not obey the single-temperature Maxwell distribution. To improve the yields of electrons in aiming energy range, further investigations are urgently demanded. Hence, various strategies have been raised to modify the structure of super-hot electron spectra and improve the energy coupling efficiency. 
	
	One of the strategies to improve optical injection of super-hot electrons is introducing the second laser pulse. For example, a counter-propagating additional laser pulse can result in an increase on injected electron energy and number \cite{Wang2008}, or excite beat wave for generating mono-energetic electron beam \cite{Faure2006}. An orthogonal beam can also dephase the electrons trapped in the wake-field and inspire transverse emittance in the accelerated bunch \cite{Dodd2004}. For solid target, limited by the laser arrangement, the coaxial schemes \cite{Chitgar2020,Zeng2015,Yu2014} and small-angle configurations \cite{Morace2019,Zhou2022} are more applicable.	To improve the electron beam, the short-wavelength pulse, whose dimensionless field strength is smaller with the same intensity ($a\propto I\lambda^{-2}$), is also widely used to generate a softer energy spectrum. The short-wavelength pulse, corresponding to a higher critical density, can also be applied to cavitate ponderomotive channels in over-dense region when the pulse duration is long enough, guiding the long-wavelength pulse into these micro structures and obtaining a more sufficient LPI. Moreover, due to the lower diffraction limit, the waist of laser can be narrowed further, reaching a higher concentration of laser energy \cite{Scott2013}. Apart from that, using short-wavelength pulse to generate super-hot electrons in FI can reduce the possible negative effect from the preplasma in gold cone \cite{Baton2008}. 
	
	By combining the two strategies above, i.e., using dual-color lasers, it is possible to further enhance the quality of electron beams. For example, a coaxial long-wavelength pulse followed by a short-wavelength pulse can produce a quasi-monoenergetic electron beam with low emittance \cite{Yu2014}, and a femtosecond laser with fundamental frequency and its harmonic co-transmit in mixed gas plasma can trigger period electron injection, generating multichromatic narrow energy-spread electron brunches \cite{Zeng2015}.
	
	A common method to generate a short-wavelength pulse is frequency-doubling. The conversion efficiency of nonlinear crystals can reaches $\sim 65\%$, and the rest energy remains as fundamental frequency component. Therefore, whether we should retain or filter out the fundamental frequency component after frequency-doubling in fast igintion, such as the DCI project in China, is an urgent realistic question.
	
	In this paper, we investigated the effects of simultaneous dual-color laser injection on solid thick targets and the underlying physical processes. In Section \ref{sec:PIC}, the simulation parameters and corresponding electron spectrum and energy transfer efficiency are given in detail, showing a significant yield increase of high-energy range electrons in dual-color injecting cases. Section \ref{sec:LangmuirWave} and \ref{sec:Channels} analyze the possible mechanisms of dual-color effects with different pulse duration. Finally, the polarization and frequency relationship are discussed in Section \ref{sec:Conclusion}, followed by the conclusion.
	
	\section{\label{sec:PIC}2D Particle-in-Cell Simulation}
	
	We assume that the maximum approachable intensity is $I_0=5\times 10^{19}$ W/cm$^2$, and the 2nd harmonic ($2\omega$) conversion efficiency is 65\% (for typical nonlinear crystal like BBO, this value can reach up to 60\textemdash70\%). Part of fundamental frequency component is retained here to investigate the influence on laser-plasma coupling and electron energy spectra in dual-color injecting cases.
	
	\subsection{\label{sec:Parameters}Simulation Parameters}
	2D PIC simulations are performed with the KLAPS code\cite{Wang2015}. The laser beams in our simulations transmit along $+x$ axis. Every beam has an intensity profile of $I(y,t)=I_\alpha e^{-y^2/\sigma^2}\sin^2(\pi t/\tau),\alpha =1,2$, with p-polarization unless otherwise specified. The footnote $\alpha$ denotes parameters of different laser beams in a single simulation case. In this work, we specify Laser 1 as the fundamental frequency beam, where $\lambda_1= 1$ \textmu m, and Laser 2 as the $2\omega$-beam. All the related parameters, for instance, the critical density $n_{cr}=\varepsilon_0 m_e \omega^2/c^2$ and dimensionless strength ${\bf a}=e{\bf E}/m_e c\omega$, are calculated with the frequency of the first pulse. The target is set to be collisionless Au plasma with a static ionization of $Z=10$. The initial density profile $n_0(x,y)=n_0(x)$ is shown as Eq. (\ref{eq:denprof}), which has an exponential pre-plasma due to the ASE pedestal.
	\begin{align}
		n_0(x)= 
		\left\{
		\begin{aligned}
			&n_{min}\exp(\frac{x-x_0}{L}),&&x_0<x<x_1\\
			&n_{max},&&x_1<x<x_2\\
			&0,&&\text{Others}
		\end{aligned}
		\right.
		\label{eq:denprof}
	\end{align}
	The maximum density, left plasma-vacuum boundary and laser waist are fixed at $n_{max}=100n_c$, $x_0=3$ \textmu m and $\sigma=10$ \textmu m. To hold the continuity of density, we take $x_1=x_0+L\ln(n_{max}/n_{min})$.
	
	To systematically investigate the effects of dual-color injection with different pulse duration, we set $\tau=1 \text{ ps } (300 T_0), 0.2 \text{ ps } (60 T_0)$ and $66.7 \text{ fs } (20 T_0)$ for different cases, where $T_0$ is the field period of the fundamental frequency laser. The $\tau$s above correspond to typical picosecond, sub-picosecond and femtosecond laser duration separately. Considering that the prepulse from a long pulse is usually stronger than a short one, the simulation box of the picosecond case is $64\times48$ \textmu m ($2048\times1536$ cells), with $L=8$ \textmu m, $n_{min}=0.1n_c$, and $x_2=63$ \textmu m. For the sub-picosecond and femtosecond cases, the simulation box is $40\times32$ \textmu m ($1280\times1024$ cells), with $L=4$ \textmu m, $n_{min}=0.01n_c$, and $x_2=39$ \textmu m. It should be note that scale length variation cause by the change of laser wavelength \cite{Scott2013} is not considered here.
	
	Detecting surfaces perpendicular to $x$ axis are placed at regular intervals of $\Delta x=2$ \textmu m to record the energy flux of the super-hot electrons which pass through the surface with $p_x>0, E_k>50$ keV for the first time. Then, the spectra of electrons recorded by the detecting surface near $x_1$ (located at $x=57$ \textmu m or $x=31$ \textmu m) are given in Section \ref{sec:Results}.
	\subsection{\label{sec:Results}Simulation Results}
\begin{figure}
		\includegraphics[width=1.0\textwidth]{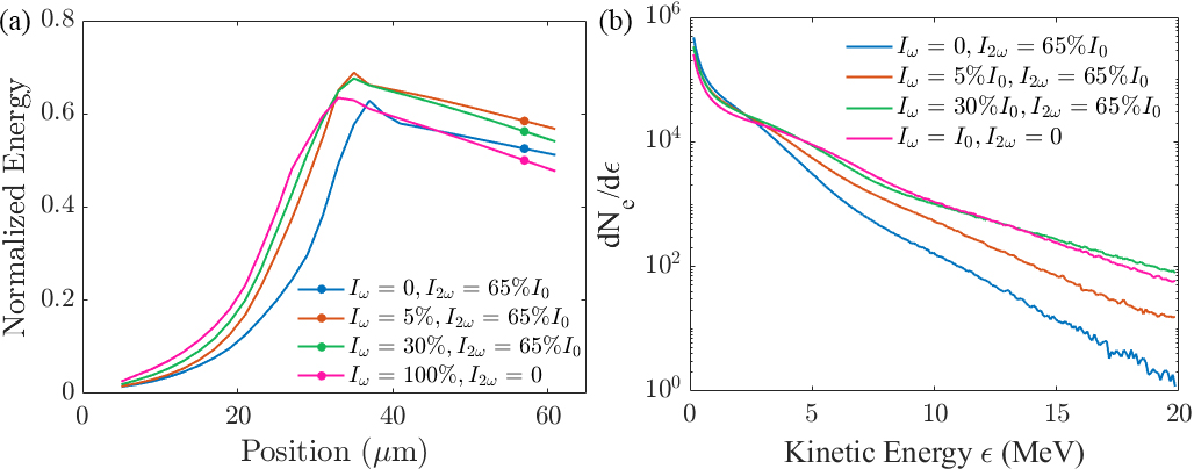}
		\caption{\label{fig:ps_spec} (a) Super-hot electron flux passing through detecting surfaces in different injecting cases. The data have been normalized with the incident total laser energy. (b) The energy spectra of super-hot electrons recorded by the surface at $x=57$ \textmu m ($n_0=85.4n_c$), corresponding to the dots marked in Figure 1(a).}
	\end{figure}

	The super-hot electron energy flux and spectrum in different injecting cases are shown in Fig. \ref{fig:ps_spec}, where $I_1=I_\omega,I_2=I_{2\omega}$. Figure \ref{fig:ps_spec}(a) infers that the dual-color injecting strategy can lift the energy coupling efficiency. A 6\% increase can be found at the detecting surface near $n_0=100n_c$ with $5\% I_0$ fundamental frequency laser doping. When the fundamental frequency component is mixed in the $2\omega$-beam, the high-energy part of electron energy spectrum has a significant rise, while the low-energy electron number falls down a little. In the case of $I_\omega=30\% I_0, I_{2\omega}=65\% I_0$, the spectrum curve has become very close to the fundamental frequency case in the high-energy range. Actually, the two curves intersect at 4.6 MeV and 12.2 MeV, which means that in both low- and high-energy regions, the dual-color injecting scheme performs better than fundamental frequency case even though there is an energy loss of 5\% in consideration of the frequency-doubling process.
	
\begin{figure}
		\includegraphics[width=1.0\textwidth]{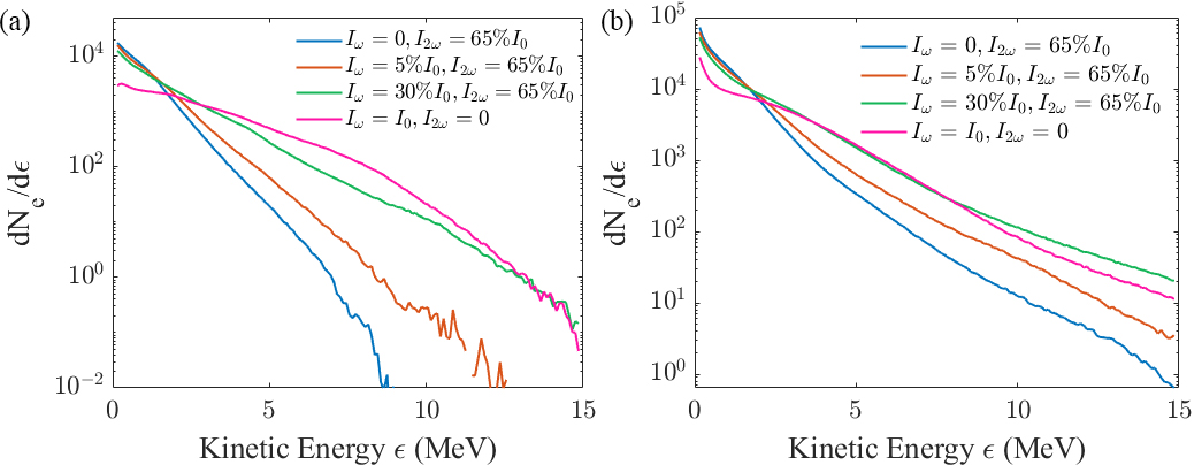}
		\caption{\label{fig:short_spec} The energy spectra of super-hot electrons recorded by the surface at $x=39$ \textmu m ($n_0=81.0n_c$), with the laser duration (a) $\tau=66.7$ fs, (b) $\tau=0.2$ ps}
	\end{figure}

	Furthermore, we investigate the electron spectra in cases of shorter pulse duration, and obtain the similar results (Fig. \ref{fig:short_spec}). The curves of dual-color injecting cases are also lifted in the high-energy range comparing to the pure $2\omega$ case for both femtosecond and sub-picosecond-laser injection. It should be pointed out that $I_\omega=30\% I_0, I_{2\omega}=65\% I_0$ case behaves better than or almost the same as fundamental frequency case over the whole energy range for sub-picosecond-laser injection [Fig. \ref{fig:short_spec}(b)]. 
	
	The mechanism of dual-color injecting effect is going to be analyzed in detail in the following two sections. Nevertheless, a caveat must be addressed in advance. Although the lower-frequency beam is usually expected to generate super-hot electrons with higher kinetic energy, the intensity also plays a part. According to the scaling estimation Eq. (\ref{for:temp}), the ponderomotive temperature is related to $I\lambda^2$, which means that it is improper to consider the two beams' contribution separately and add the electrons produced by every single laser together linearly. Taking the $I_1=5\% I_0, I_2=65\% I_0$ case as a counter-example, $(I_1\lambda^2_1)/(I_2\lambda^2_2)=0.31<1$, the fundamental frequency laser corresponds to a even lower "ponderomotive temperature", while the high-energy part of the spectrum curve soars compare to the $2\omega$-beam injection case. In a word, investigations of the associative effects of co-transmission of dual-color beams in plasma are inevitably required.
	
	\section{\label{sec:LangmuirWave}Longitudinal Electric Field Excited by the Dual Wavelength Laser Field}
\begin{figure}
		\includegraphics[width=1.0\textwidth]{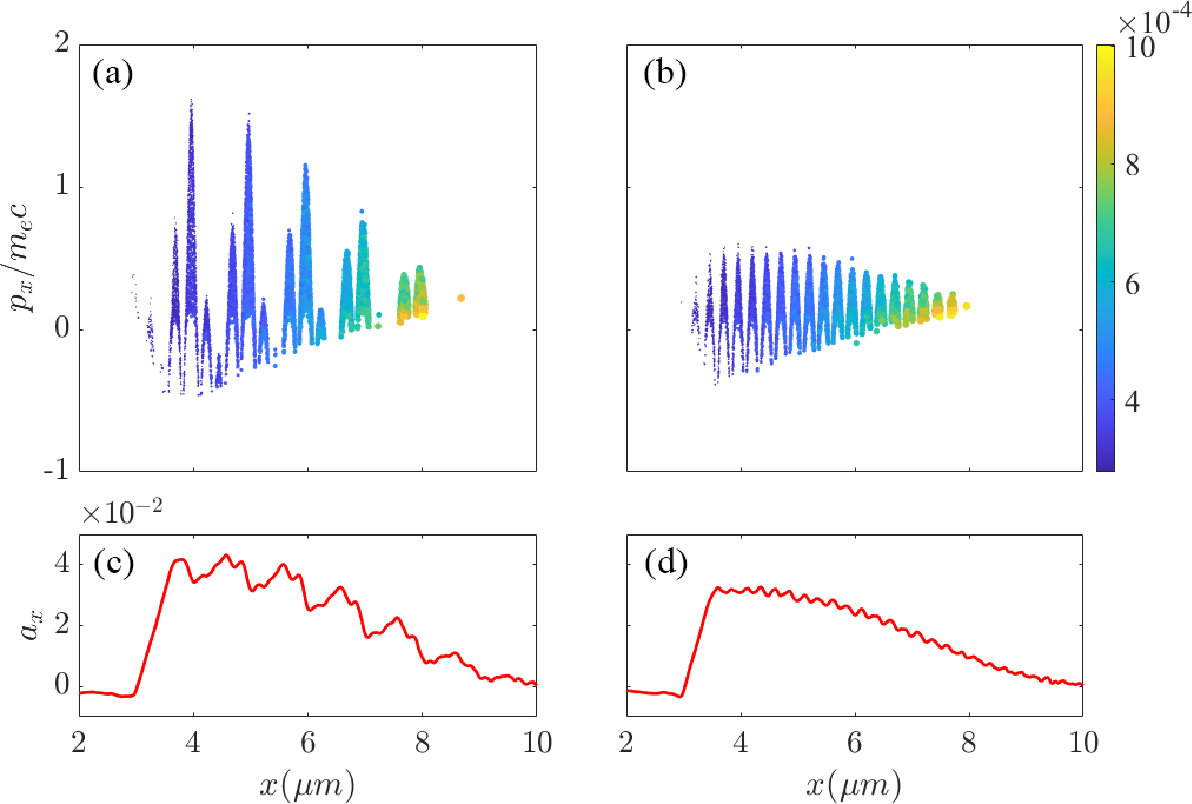}
		\caption{\label{fig:phase_ex} The phase space diagram ($p_x-x$) and dimensionless field strength $a_x$ on $x$ axis of (a), (c)  $I_1=5\%, I_2=65\% I_0$ case and (b), (d) $I_1=0, I_2=65\% I_0$ case at $t=33.3$ fs ($10T_0$). The data have been normalized with the incident total laser energy. Figure (a) and (b) plot 0.5\% of the electrons randomly whose kinetic energy are larger than 50 keV, with a shared coordinate axis. The size and color of points denote the weight of picked simulation particles.} 
	\end{figure}

	Since the increase of high-energy part of the electron flux appears in cases with various pulse duration scale, it is deduced that mixture of the two beams would alter the electron acceleration process from the linear stage of laser-plasma interaction. The phase space diagram and dimensionless field strength $a_x$ on $x$ axis of two different femtosecond-laser injecting schemes, $I_1=5\%, I_2=65\% I_0$ (dual-wavelength beams) and $I_1=0, I_2=65\% I_0$ (pure $2\omega$ beam), are compared in Fig. \ref{fig:phase_ex}.
	
	With a $5\% I_0$ mixture of $2\omega$ laser, the phase space diagram deviates from a modulated structure symmetric about $p_x=0$ [Fig. \ref{fig:phase_ex}(b)]. Several peaks and sub-peaks emerge with certain space period ($\sim \lambda_1$), which correspond to electrons with much higher forward momentum and kinetic energy. Changes are also found in longitudinal electric field diagrams. Sawtooth structure appears on $a_x$ plot shown in Fig. \ref{fig:phase_ex}(c), with a matched spacial period as peaks in phase space diagram. When the intense laser field interacts with underdense plasma, a longitudinal wakefield would be exited behind the laser front, which can trap electrons and accelerate them as the wave propagates \cite{Tajima1979,Brantov2008}. The quasi-static approximation (QSA) can be applied to plane wave propagating in uniform underdense cold plasma, until the wave envelope changes significantly, whose characteristic timescale is usually the Rayleigh diffraction time \cite{Gibbon2005}. In this scheme, there is a coordinate transformation $\xi=x-v_pt,\tau=t$ in 1D problem and the longitudinal field $a_x(\xi)=-\dd{\phi}/\dd{(k_p\xi)}$ satisfies Poisson's equation \cite{Dalla1993}:
	\begin{equation}
		\dv[2]{\phi}{\xi}=k_p^2\gamma^2\left[
		\frac{\beta_p(1+\phi)}{\sqrt{(1+\phi)^2-\gamma^{-2}(1+a^2)}}-1
		\right],
		\label{for:Poisson_vp}
	\end{equation}
	where $\beta_p\sim 1$ is the phase velocity of $a_x$ excited by the laser field, $\gamma=1/\sqrt{1-\beta_p^2}$ and $k_p=\omega_p/c$. Particularly, in rarefied plasma ($n\ll n_c$), $\beta_p=1$ and Eq. (\ref{for:Poisson_vp}) can be reduced to Eq. (\ref{for:Poisson_c}) \cite{Bulanov1989}:
	\begin{equation}
		\dv[2]{\phi}{\xi}=k_p^2
		\frac{(1+a^2)-(1+\phi)^2}{2(1+\phi)^2}
		.
		\label{for:Poisson_c}
	\end{equation}
	Once the pulse envelope is substituted into the Poisson's equation, the well-known wakefield solution can be obtained numerically \cite{Bulanov1989,Dalla1993}, which is theoretical basis of wakefield acceleration. Here, since the envelopes of laser beams are the same, the oscillating term of field must be taken into consideration. For the validity of QSA, it is assumed that both of the two-color beams have the same phase velocity $c$, suggesting that the dispersion is ignorable and $\omega_1/\omega_2=k_1/k_2$.
	
\begin{figure}
		\includegraphics[width=1.0\textwidth]{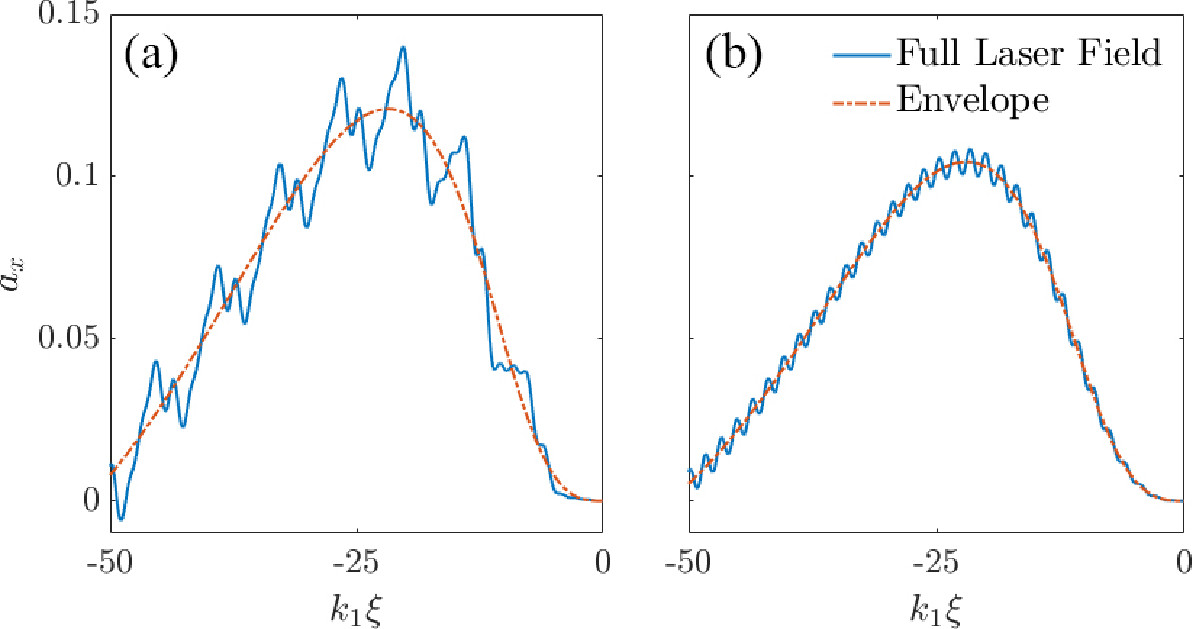}
		\caption{\label{fig:analytic} The numerical solutions of Eq. \ref{for:Poisson_c} of (a) dual-color case ($a_1=2a_2=5$), and (b) pure-$2\omega$ case ($a_2=0$). Both of the two cases take $\hat{\bf e}_1=\hat{\bf e}_2=\hat{\bf e}_y$, $k_2/k_1=2$, $n_c/n=25$, and $N=20$. The solid lines denote that $a$ takes the full version of Eq. (\ref{for:field}), and the dashed lines correspond to solutions derived from a laser field envelope.}
	\end{figure}

	When we normalize $k_1\xi\rightarrow\xi$, and take
	\begin{equation}
		{\bf a}(\xi)=\left\{
		\begin{aligned}
			&\sin \left(\frac{\xi}{2N}\right)
			\left[
			a_1\hat{\bf e}_1 \sin(\xi) + a_2\hat{\bf e}_2 \sin\left(\frac{k_2}{k_1}\xi\right)
			\right], &&0<\xi<2\pi N\\
			&0, &&Others
		\end{aligned}
		\right.,
		\label{for:field}
	\end{equation}
	where $N=\tau/T_1$ is the pulse duration normalized by field circulation of Laser 1, the longitudinal field can be solved numerically. Figure \ref{fig:analytic} compares $a_x$s of the dual-color (a) and pure-$2\omega$ injecting case given by Eq. (\ref{for:Poisson_c}). The consideration of full laser field brings fluctuations on the smooth $a_x$ curve (shown as the dashed line in Fig. \ref{fig:analytic}), which matches the PIC simulation result of Fig. \ref{fig:phase_ex}(d). In the dual-color case, The sawtooth structure is observed clearly, leading to the deduction that co-transmission of dual-color transverse field can modulate the longitudinal static electric field, forming periodic sawtooth structure. The exited positive longitudinal field would hamper forward acceleration. However, for electrons located at the troughs of modulated wave, they suffer a decreased damping force and can gain a higher forward momentum, corresponding to the peaks in Fig. \ref{fig:phase_ex}(a). The modulating strength is stronger with a mixture of higher amplitude fundamental frequency beam. Therefore, the high-energy part of electron energy spectrum curve is lifted, resulting in harder spectrum and a higher conversion efficiency. 
	
	In longer duration cases, the mechanism discussed above is also valid. However, since the intensity grows much slower than ultra-short pulse, these electrons with relatively higher initial forward momentum at the wave troughs would be further accelerated by subsequent laser field directly or other mechanisms like stochastic acceleration \cite{Sheng2002,Meyer-ter-Vehn2007}.
	
	\section{\label{sec:Channels}Long Pulse: Microchannels Forming on the Critical Surface}
	Although modulation of $a_x$ can generate a part of electrons as seed for later acceleration, the longitudinal field far behind the wavefront, where the laser amplitude actually rises in long pulse cases, would become chaotic, leading to the destruction of sawtooth structure. Besides, there is enough time for plasma to respond the perturbation from laser fields and for nonlinear effects to develop. One of the direct consequences is that a set of microchannels is cavitated due to the perturbation of electron density by ponderomotive force and transverse instabilities \cite{Huang2013,Wan2018}. Laser fields can propagate along these microchannels \cite{Esarey1993,Pukhov1996}, and thus a longer distance for LPI is approachable, resulting in gains on energy coupling efficiency. The acceleration of electrons in these channels would also be affected by the static fields, where some of electrons are expelled out of the channel center, forming a static electron field $E_y^s\propto y$ and $B_z^s\propto -y$ \cite{Pukhov1999,Arefiev2012}.
	
\begin{figure}
		\includegraphics[width=0.8\textwidth]{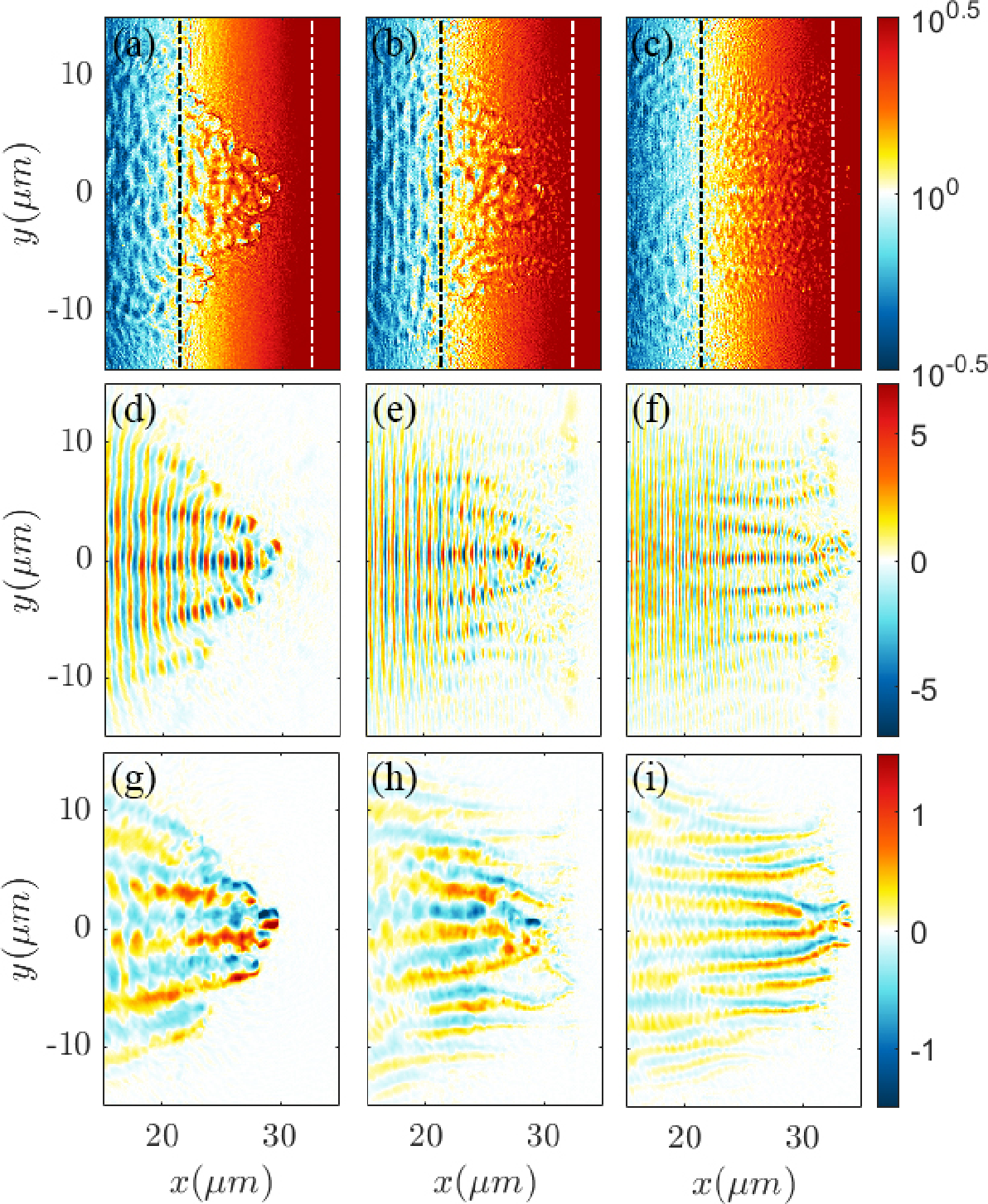}
		\caption{\label{fig:fields} Diagrams of (a)-(c) electron density, (d)-(f) normalized electric field $a_y$, and (g)-(i) normalized static magnetic field $\langle b_z \rangle$ in different picosecond simulation cases at $t=400$ fs. The first row: $I_1=100\% I_0, I_2=0$ (pure fundamental frequency); the second row: $I_1=30\% I_0, I_2=65\% I_0$ (dual-color); the second row: $I_1=0, I_2=65\% I_0$ (pure $2\omega$). The black and white dash-dotted lines in (a)-(c) correspond to $n_0(x)=n_{cr}$ and $n_0(x)=4n_{cr}$ (critical surface of $2\omega$ beam) separately.}
	\end{figure}

	Figure \ref{fig:fields} shows the electron density, normalized electric field $a_y$ and normalized static magnetic field $\langle b_z \rangle$ in different picosecond simulation cases at $t=400$ fs. The magnetic fields are averaged over two laser field periods. To show the details of laser-plasma interaction region, the plot spacial range is limited to $X\times Y\in[15,35]\times[-15,15]$ (\textmu m). It is confirmed with the density diagrams that the over-dense region where $n_0>n_{cr}$ is perturbed, displaying some certain parallel structures. There are also channel structures in electric field plots which overlap the perturbed area in density diagrams, implying that laser fields do propagate along the microchannels. Hence, super-hot electrons can be generated and transport in these channels towards the rear surface of the target, which is consistent with the static magnetic field configuration in Fig. \ref{fig:fields}(g)-(i). It should be noticed that the width of microchannels varies among different cases, which could be related to parameters such as the normalized field strength, plasma density and scale length. Obviously, a longer and wider channel is beneficial to LPI and energy coupling, which can be considered as one of the reasons that dual-color injecting scheme has a higher coupling efficiency.
	
	Although plasma in microchannels is perturbed and expelled by the laser field, the fundamental frequency component can still penetrate into the over-dense region. In the pure fundamental frequency case, one can ascribe this phenomenon to the well-known relativistic transparency (RT) effect \cite{Kaw1970}. Since the zeroth-order dispersion relationship for relativistic laser field is $\omega^2=k^2c^2+\omega_p^2/\gamma_0, \gamma_0=\sqrt{1+\langle a^2\rangle/2}$, the effective plasma frequency decreases as the rise of normalized field amplitude, making plasma transparent for the electromagnetic wave. 
	
\begin{figure}
		\includegraphics[width=0.5\textwidth]{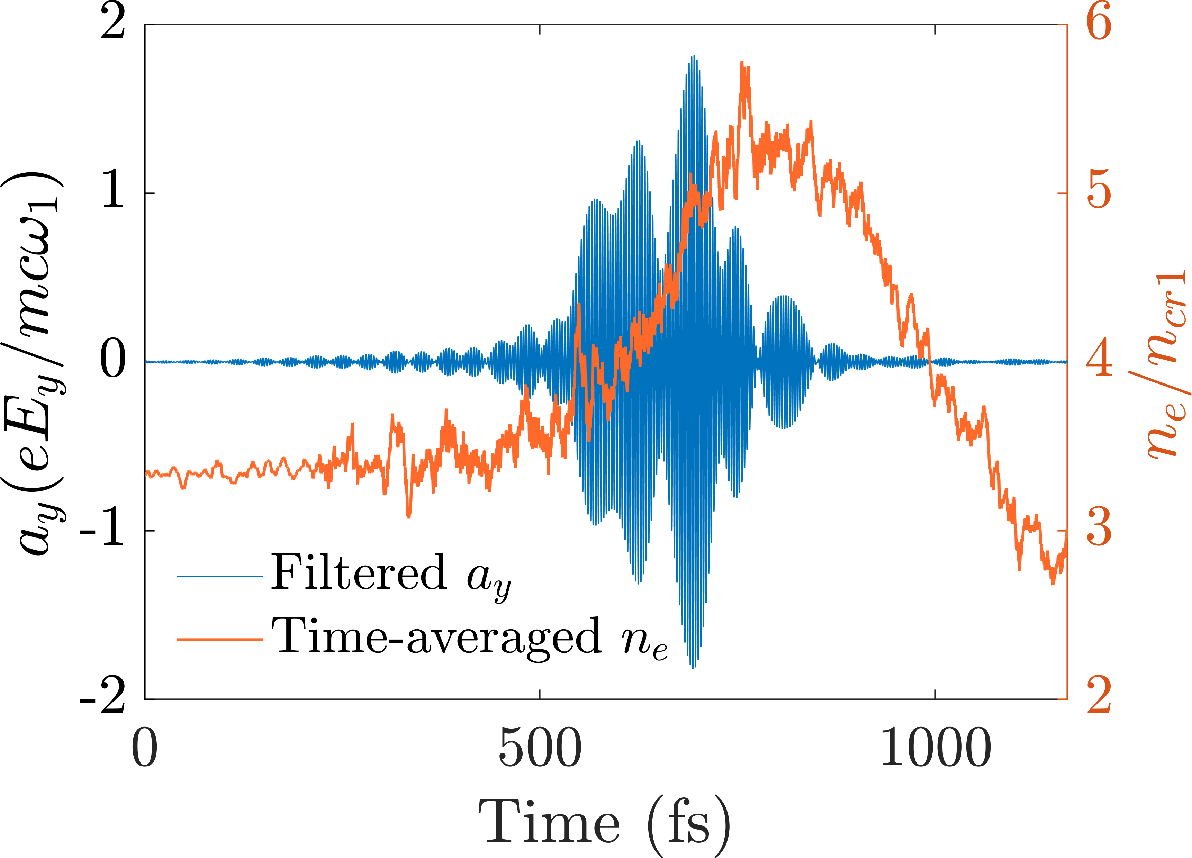}
		\caption{\label{fig:density vs ay} Smoothed electron density $n_e$ and normalized laser fields $a_y$ filtered within $[0.95\omega_0,1.05\omega_0]$ at $(x,y)=$(31 \textmu m, 0) in case $I_\omega=5\% I_0, I_{2\omega}=65\% I_0$. Since the cavitation process would push the plasma forward, the local density rises from $3.32n_{cr1}$ to $5.78n_{cr1}$ in $750$ fs, then falls down to $2.73n_{cr1}$.}
	\end{figure}

	While in the dual-color injecting case, even though the intensity of fundamental frequency component is as low as $5\% I_0$, it can be detected where the local electron density rises up to several times more than $n_{cr1}$. Figure \ref{fig:density vs ay} shows the time-dependent evolution of smoothed $n_e$ and filtered normalized electric field $a_y$ at $(x,y)=$(31 \textmu m, 0). Between 500 fs to 800 fs, strong fundamental frequency signals can be detected even if the local $n_e$ is much higher than the RT-modified critical density $\gamma_{1}n_{cr1}=1.63n_{cr1}$, which can be attributed to the relativistic electromagnetically induced transparency (EIT) mechanism \cite{Tie-huaiZhang}. In the physical picture of EIT, when a low-frequency laser $(a_1,\omega_1)$ co-transmit with a pump laser with $\omega_2>\omega_p$ in over-dense plasma, the induced current of Laser 1 tends to be canceled by the beat current of pump wave and plasma wave, leading to the transparency of Laser 1 \cite{Harris1996,Gordon2000a,Gordon2000b}.
	
	At the late stage of LPI, the microchannels would expand transversely and break down. Eventually, these channels merge and forming a cavity with the radius $r\sim \sigma$, which can be converted to the typical hole-boring picture \cite{Wilks1992,Pukhov1997}. However, the laser intensity is not extremely strong and pulse duration is relatively short in this paper, the subsequent hole-boring process is weak and can be omitted here \cite{Kemp2008,Weng2012}.	
	
	\section{\label{sec:Conclusion}Dependency on polarization and frequency relationship}
\begin{figure}
		\includegraphics[width=1\textwidth]{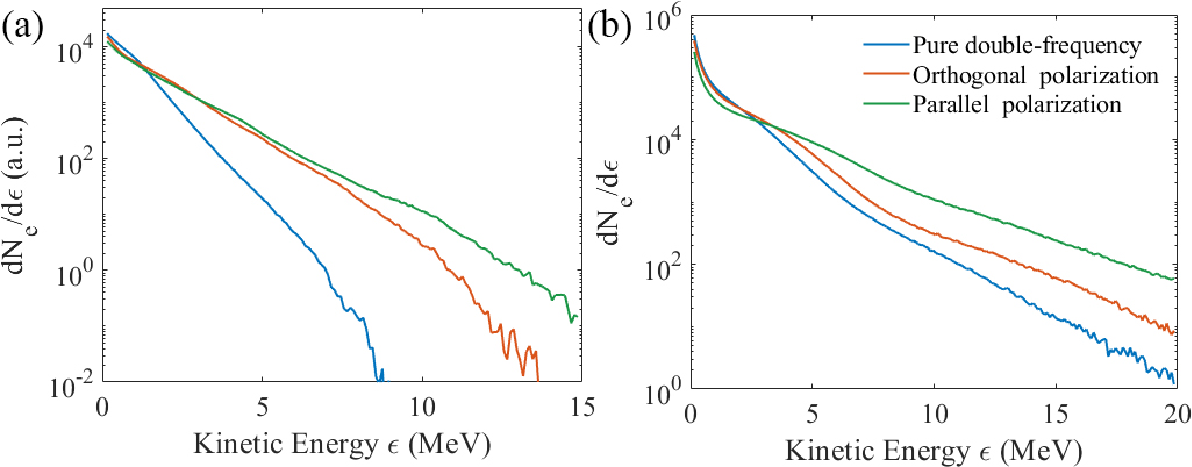}
		\caption{\label{fig:polarization} The energy spectrum of (a) femtosecond laser and (b) picosecond laser simulation. The peak laser intensity in pure $2\omega$ injecting scheme is $I_1=0, I_2=65\%I_0$. The polarization direction of Laser 1 in orthogonal and parallel polarization cases are $\hat{\bf e}_1=\hat{\bf e}_z$ and $\hat{\bf e}_1=\hat{\bf e}_y$ separately, where the direction of Laser 2 is kept at $\hat{\bf e}_2=\hat{\bf e}_y$.}
	\end{figure}

	In Section \ref{sec:PIC}, \ref{sec:LangmuirWave} and \ref{sec:Channels}, both of the two beams are fixed at p-polarization. In some cases, the nonlinear crystals can generate second harmonic waves with the polarization direction perpendicular to the pump beam. To investigate the effect of beam polarization on electron energy spectrum, we changed the polarization direction of Laser 1 to $\hat{\bf e}_1=\hat{\bf e}_z$ (s-polarization) and took the energy proportion $I_1=30\%I_0, I_2=65\%I_0$ in dual-color injecting scheme, showing the electron energy spectrum of femtosecond and picosecond cases in Figure \ref{fig:polarization}. The orthogonal polarization directions have a negative effect on the super-hot electron generation in both of the two cases, especially in high-energy region. It is unsurprising since one can calculate the modulation of longitudinal field is greatly reduced when $\hat{\bf e}_1\cdot \hat{\bf e}_2 = 0$. According to Eq. (\ref{for:field}) and letting $a_1=\epsilon a_2$, $a^2={\bf a}\cdot{\bf a}=\sin^2(k_2\xi/k_1)+(\hat{\bf e}_1\cdot \hat{\bf e}_2)\epsilon \sin(\xi)\sin(k_2\xi/k_1)+\epsilon^2 \sin^2(\xi)$, and $\hat{\bf e}_1\cdot \hat{\bf e}_2 = 0$ leads to the first-order perturbation term being zero. In long-pulse injecting case, the microchannels still exist, where an external driving term appears in parametric instability equations \cite{Arefiev2012} due to the orthogonal polarization. Since the induced currents of the two laser beams fluctuate in different directions, the EIT effect is expected to vanish \cite{Tie-huaiZhang}, resulting in the decrease of super-hot electron output. However, it should be noted that the simulations here are in 2D geometry, where the inhomogeneous effects along $z$ direction cannot appear and need further investigation of 3D simulations.
	
\begin{figure}
		\includegraphics[width=1\textwidth]{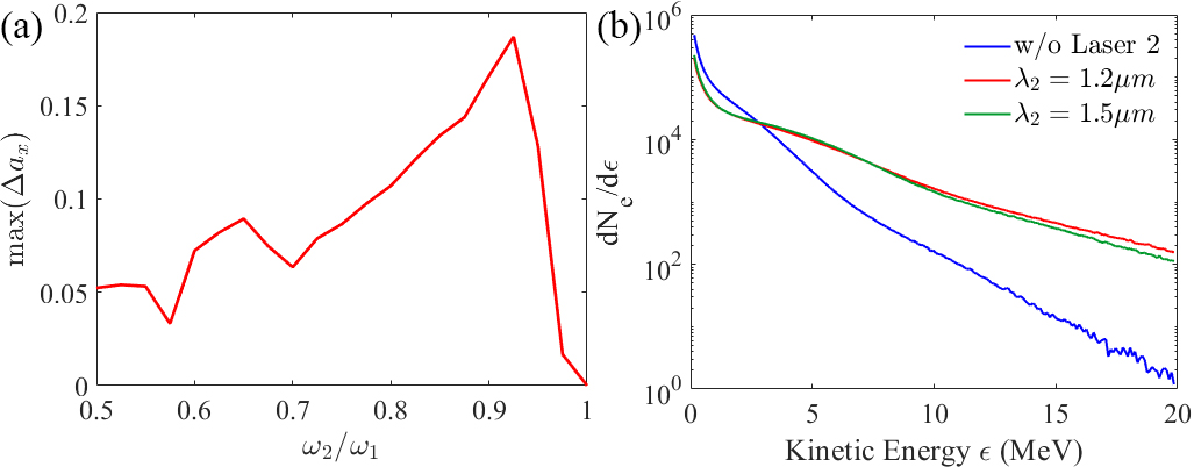}
		\caption{\label{fig:frequency} (a) The maximum of $\Delta a_x=a_{x,env}-a_{x,full}$ within $k_1\xi\in[-50,0]$ calculated from Eq. (\ref{for:Poisson_c}) where $2a_1=a_2=5$ in Eq. (\ref{for:field}). (b) Energy spectrum of different injecting schemes, where $I_1=I_0, I_2 = 10\% I_0$ and $\lambda_1=1$ \textmu m are fixed. The red and green lines denote $\lambda_2=1.2$ \textmu m ($\omega_2/\omega_1=5/6$) and $\lambda_2=1.5$ \textmu m ($\omega_2/\omega_1=2/3$) separately.}
	\end{figure}

	According to Eq. (\ref{for:Poisson_c}), the strict harmonic relationship in dual-color effects discussed in this article is not necessary, except that EIT mechanism requires $\omega_p<\sqrt{\gamma}\omega_2<2\omega_p, \omega_2-\omega_1<\omega_p/\sqrt{\gamma}$ \cite{Tie-huaiZhang}. We calculated the maximum of $\Delta a_x=a_{x,env}-a_{x,full}$ in Fig. \ref{fig:frequency}(a), which is the difference between longitudinal fields excited by the full field and laser envelope in Eq. (\ref{for:Poisson_c}). The frequency of Laser 2 varies from $\omega_2=\omega_1/2$ to $\omega_2=\omega_1$. $(\Delta a_x)_{max}$ is presented to evaluate the strength of longitudinal field modulation, and the results show that the perturbation also takes place in anharmonic cases. Figure \ref{fig:frequency}(b) shows that a $10\%$ mixture of long-wavelength pulse in picosecond injecting cases can lifting the high-energy part of spectrum significantly, even though the frequency of Laser 2 is not an integral multiple of Laser 1. The high tolerance of frequency relationship indicates that it could be possible to obtain an extreme high-energy electron beam by injecting an additional low-intensity but long-wavelength beam along with the main pulse.
	
	\section{\label{sec:Summary}Conclusion}
	In summary, we have shown that the reservation of fundamental frequency component can evidently lift the high-energy part of electron spectrum and enhance the laser-plasma coupling process with different pulse duration (from femtosecond to picosecond). The dual-color mixing effect are investigated and attributed to the modulation of longitudinal field and the formation of microchannel where different physical mechanisms such as relativistic EIT would emerge with the second beam's injection. The polarization direction and frequency relationship required are also discussed. The results in this paper can be applied to fast ignition schemes (for example, DCI) and experimental designs of super-hot electron generation with different pulse durations.
	
	\begin{acknowledgments}
		This work was supported by the Strategic Priority Research Program of Chinese Academy of Sciences (Grant Nos. XDA25050300, XDA25010300, XDA25010100), the National Key R\&D Program of China (Grant No. 2018YFA0404801), and National Natural Science Foundation of China (Grant No. 11827807 and 11775302), and the Fundamental Research Funds for the Central Universities, the Research Funds of Renmin University of China (20XNLG01). Computational resources have been partially provided by the Physical Laboratory of High Performance Computing at Renmin University of China.
	\end{acknowledgments}
	
	\bibliographystyle{apsrev4-1} 
\end{document}